\documentclass[prb,twocolumn,amssymb] {revtex4}
\usepackage{graphicx}
\begin{document}
\title{Magnetically Stabilized Order II:
\\ Critical states and algebraically ordered nematic spin liquids in one-dimensional
optical lattices}
\author{ Hui Zhai$^{\dagger}$ and
Fei Zhou$^{\dagger\dagger}$} \affiliation{$^{\dagger}$ Center for
Advanced Study, Tsinghua University, Beijing, People's Republic of China}
\affiliation{$^{\dagger\dagger}$ Department of Physics and
Astronomy, University of British Columbia,\\ 6224 Agricultural
Road, Vancouver, British Columbia, Canada, V6T 1Z1}
\date{\today}
\begin{abstract}
We investigate the Zeeman-field-driven quantum phase transitions
between singlet spin liquids and algebraically ordered $O(2)$
nematic spin liquids of spin-one bosons in one-dimensional optical
lattices. We find that the critical behavior is characterized by
condensation of hardcore bosons instead of ideal magnons in high-
dimensional lattices. Critical exponents are strongly renormalized
by hardcore interactions and critical states are equivalent to the
free Fermion model up to the Friedel oscillations. We also find
that the algebraically ordered nematic spin liquids close to
critical points are fully characterized by the Luttinger-liquid
dynamics with Luttinger-liquid parameters magnetically
tunable. The Bethe Ansatz solution has been applied to determine
the critical magnetization and nematic correlations.
\\ PACS number: 03.75.Mn, 05.30.Jp, 75.10.Jm.
\end{abstract}
\maketitle

%\begin{multicols}{2}

%\narrowtext

\section{Introduction}

It has been well known that spontaneous (continous) symmetry breaking
states usually support gapless Goldstone modes which give arise
various quantum fluctuations of local order
parameters\cite{Ma79,Zinn-Justin93,Chaikin95,Sachdev99}.
In one dimension at zero temperature, long wave length
quantum fluctuations can be overwhelming and extremely harmful to
the establishment of long range order. Depending on details of
energetics and topology of order parameter manifolds, quantum
fluctuations might result in complete symmetry restoring by
driving the states away from the spontaneous symmetry breaking
solutions, or lead to algebraically long range order instead of
the conventional off-diagonal long range order in
high-dimensional systems.

For $O(3)$ nematic states of spin-one bosons in optical lattices
(either condensates or Mott states) studied recently in quite a
few
articles\cite{Demler02,Zhou01,Zhou03a,Zhou03b,Snoek04,Imambekov03,Imambekov04,Zhou04},
indeed in one-dimensional lattices the symmetry is completely
restored and resultant states are rotationally invariant. Unlike
high-dimensional nematic states, spin excitations in these
states are fully gapped as a result of symmetry restoring.

From the point of view of $(1+1)d$ Euclidean space, one can also
attribute, at least partially, the disruption of long range order
to the proliferation of Meron- or Skyrmion-like topological
instantons. This fact suggests that the topology of vacuum
manifold (mean field) should play an important role in maintaining or
destroying long range order.

For instance, when a magnetic field is applied to $O(3)$ nematic
states, the easy axis of the nematic order parameter tensor is
pinned in a plane perpendicular to the external field. This lows
the symmetry of nematic order and the $O(3)$ symmetry is reduced
to $O(2)$ one, i.e. the state becomes a planar nematic state. For
nematic condensates this mechanism leads to a vacuum manifold of
$[S^1\times S^1]/Z_2$ instead of $[S^1\times S^2]/Z_2$ in the
absence of magnetic fields.

The above observation then implies that in one-dimension in the presence of {\em external
magnetic fields} long wave length
fluctuations should not completely disrupt nematic ordering in a
plane perpendicular to the external fields as pointed out in a previous
article\cite{Zhou01}. Instead one should
expect an algebraically ordered nematic spin liquid. In this paper
we are going to investigate this possibility from a very different
starting point and develop a theory which fully and quantitatively
characterizes field-induced algebraically ordered nematic spin
liquids in one-dimensional optical lattices.

A closely related subject, magnetically stabilized nematic order in three-dimensional
bipartite lattices, has also been
studied recently\cite{Zhou04}.
It was found that the stabilization of nematic order can be studied
via a mapping to the ferromagnetic $XXZ$ model and the critical
phenomenon is identical to condensation of
ideal bosonic magnons. Away from critical fields the nematically ordered
states are described by weakly interacting dilute magnons.

We are going to employ
the previously developed mapping between the low energy Hilbert space
close to critical fields and
the ferromagnetic $S=1/2$ $XXZ$ model to further analyze one-dimensional
field-induced
algebraically ordered spin nematic liquids.
Though the general approach
employed below is similar to that developed for three-dimensional lattices,
the unique feature of dilute magnons in the one-dimension limit makes
the current situation distinct and physics qualitatively different from
that discussed in the previous article.
In fact, it is the distinct properties of {\em fermionized} dilute magnons
which lead to {\em anomalous} critical exponents as well as
fascinating algebraically ordered nematic liquids discussed below.

In Sect. II, we are going to review some basic results of the
mapping between the field-driven quantum phase transitions
(between spin singlet Mott states and nematic Mott states) and the
ferromagnetic $S=1/2$ $XXZ$ model. We further demonstrate the
linear relation between the pseudo-spin operators in the $XXZ$ model
and nematic tensor order parameter operators. In Sect. III, we
study the critical behavior of these quantum phase transitions in
the Holstein-Primakov boson model, and point out that
one-dimensional quantum fluctuations strongly renormalize the
critical exponents. Unlike in three-dimensional lattices where the
critical point is fully characterized by {\em ideal bosons}
(noninteracting), in one-dimensional lattices the critical point
is characterized by {\em hardcore bosons with an infinitely large
interaction constant}. And we believe our evaluation of the
critical exponent is exact.

In Sec. IV, we further investigate the critical point physics
using the Jordan-Wigner's fermionic representation. We obtain an
identical critical exponent as in Sec. III. We argue that the
critical point is also fully characterized by the free Fermion
model; however the Friedel oscillations, one of the basic aspects
of Fermions, do not appear in the correlation function of nematic
order operators.

In Sec. V, we further study the one-dimensional algebraically
ordered nematic spin liquids close to critical fields using a
strongly interacting boson model. We employ the technique of
bosonization to evaluate various correlation functions in the
nematic spin liquids and argue that in general these liquids, up
to the Friedel oscillations are equivalent to the Luttinger
liquids of Fermions. In Sec. VI, we conclude our studies and
discuss some consequences of these theoretical findings.

\section{The effective theory close to critical points: the ferromagnetic
$S=1/2$ $XXZ$ model}

We introduce $\psi^\dagger_{k\alpha}(\psi_{k\alpha})$,
$\alpha=x,y,z$ as the creation (annihilation) operators for
a spin-one particle in three orthogonal states at site $k$. The
effective Hamiltonian for Mott states of spin-one bosons with
antiferromagnetic interactions in optical lattices in an external
field can be conveniently expressed
as\cite{Zhou03a,Imambekov03,Snoek04,Imambekov04,Zhou04},

\begin{eqnarray}
&& H=E_s\sum_k \hat{\bf S}^2_k -\sum_k \hat{\bf S}_{kz} H_z\nonumber \\
&& -J_{ex}\sum_{\langle kl\rangle } (\hat{Q}_{\alpha\beta}(k) \hat{Q}_{\beta\alpha}(l)
+h.c.);\nonumber \\
&& \hat{Q}_{\alpha\beta}(k)=\psi^\dagger_{k\alpha}\psi_{k\beta}-\frac{1}{3}
\delta_{\alpha\beta}\psi^\dagger_{k\gamma}\psi_{k\gamma}.
\label{MottH}
\end{eqnarray}
Here $E_{s}$ is the "bare" spin gap studied in
Ref.(\onlinecite{Zhou03a}) and $J_{ex}$ is the strength of
exchange interaction. $\hat{\bf
S}_{k\alpha}=-i\epsilon_{\alpha\beta\gamma}
\psi^\dagger_{k\alpha}\psi_{k\beta}$ is the total spin operator
for lattice site $k$.
 $\hat{Q}_{\alpha\beta}(k)$ is a bilinear tensor
operator, the expectation of which yields the nematic order
parameter. Finally, the sum over $\langle kl\rangle $ represents
a sum over all neighboring sites. And again we are interested in
the case with an even number of particles per lattice site, though
all conclusions are qualitatively correct for an odd number of
particles per site.

In high-dimensional lattices, when the external field is absent
and $\eta=J_{ex}/E_s$ is smaller than a critical value $\eta_c$,
the ground state is a spin singlet Mott state for an even number
of particles per site ($N$). And when $\eta$ is larger than
$\eta_c$, the ground state is a nematic Mott state. The critical
value $\eta_c$ has been calculated in various limits and turns out
to be of order of unity\cite{Snoek04}.

In one-dimensional lattices, in the absence of external magnetic
fields the ground state is always a spin singlet one and the mean
field solution with nematic ordering is unstable. When $J_{ex}$ is
set to be zero, at each lattice site the lowest energy state is a spin
singlet one ($S=0$) and the first excited states are fivefold
degenerate $S=2$ states. The ground state in this case is simply
a product of spin singlets at each lattice.

When an external field is applied along the $z$-axis, one of the
five-fold degenerate $S=2$ states approaches the spin singlet and level
crossing takes place. Generally speaking
close to the level crossing points at each site, two
states $|S,S_z>$ and $|S+2, S_z+2>$ become degenerate;
the on-site low energy Hilbert space is spanned by the two states
involved in the level crossing.

It is therefore convenient to express the effective low energy
Hamiltonian in this pseudo-spin subspace. Consider the crossing
between states $|0,0>$ and $|2,2>$. If one introduces the on-site
pseudo spin Hilbert space ${\cal S}_k$ which consists of two
states,

\begin{equation}
|\! \uparrow \rangle =|S=2, S_z=2\rangle,
|\! \downarrow \rangle =|S=0, S_z=0\rangle,
\end{equation}
then the truncated Hilbert
space
for the whole lattice ${\cal S}_T$ is the product of pseudo spin Hilbert
space
${\cal S}_k$ at each site $k$, i.e.
${\cal S}_{T}={\cal S}_1\otimes {\cal S}_2\otimes...
\otimes{\cal S}_k \otimes...$.

Defining $|\! \uparrow \rangle $ and $|\! \downarrow \rangle $ to
be two eigenstates of Pauli-matrix $\sigma_z$, $\sigma_z |\!
\uparrow \rangle = |\! \uparrow \rangle$, $\sigma_z |\! \downarrow
\rangle =-|\! \downarrow \rangle$, one obtains the effective
Hamiltonian which is represented by the ferromagnetic $XXZ$ model
in an effective external field along the
$z$-direction\cite{Zhou04},

\begin{equation}
\frac{H_{XXZ}}{\epsilon_0 J_{ex}}=- \sum_{\langle kl\rangle }
\sigma_{k\alpha}\sigma_{l\alpha}
-\beta\sum_{\langle kl\rangle }\sigma_{kz}\sigma_{lz}-h_z
\sum_k \sigma_{kz}.
\label{XXZ1}
\end{equation}
Eq.(\ref{XXZ1}) holds when the truncation is valid, which requires
that $J_{ex}$ should be much less than $E_s$. Here$\epsilon_0$,
$\beta$ and $h_z$ were calculated in Ref.(\onlinecite{Zhou04}).

The Hamiltonian in Eq.(\ref{XXZ1}) has an $O(2)$ invariance in the
$xy$-plane
corresponding to the $O(2)$ rotational invariance in
the microscopic Hamiltonian in Eq.(\ref{MottH}) when
the external magnetic fields are applied along the $z$-direction.
This $O(2)$ invariance represents the $O(2)$ nematic symmetry we are going to
examine. Furthermore,
in general the truncation can be applied in the vicinities of all
critical
points where level crossings between $|S, S_z=S\rangle $ and $|S+2, S_z=S+2\rangle $ occur,
$S+2 \leq N$.

The main issue dealt in this work is the induced
spontaneous symmetry breaking in the presence of external magnetic
fields. Particularly we are interested in the breaking of $O(2)$
rotational symmetry in the $xy$-plane when a field is applied
along the $z$-axis. Unfortunately all operators that are linear in
$\psi^\dagger_\alpha$, or $\psi_\alpha$ vanish identically in Mott
states regardless whether the rotational symmetry is broken or
not; so linear operators fail to distinguish two phases which
interest us.

The rank-two-tensor order parameter introduced on the other hand
vanishes only in rotationally invariant spin singlet states and
develops nontrivial structures once states break the $O(2)$
symmetry. For this reason we introduce the projected nematic order
parameter $\hat{Q}^P_{\alpha\beta}$($\alpha,\beta=x,y,z$) which is
designed to signify the $O(2)$ symmetry breaking in the $xy$ plane
as follows
\begin{equation}
\hat{Q}^P_{\alpha\beta}=\hat{Q}_{\alpha\beta}-
(\hat{Q}_{\alpha'\beta'}\Pi^1_{\beta'\alpha'}) \Pi^1_{\alpha\beta}
-(\hat{Q}_{\alpha'\beta'}\Pi^2_{\beta'\alpha'})\Pi^2_{\alpha\beta}.
\label{PNOP}
\end{equation}
Two constant tensors we would like to project away are defined
as\cite{Zhou04}

\begin{eqnarray}
\Pi^1_{\alpha\beta}=\frac{1}{\sqrt{2}}
i\epsilon_{\alpha\beta\gamma} {\bf s}_\gamma,
\Pi^2_{\alpha\beta}=\frac{3}{\sqrt{6}} ({\bf s}_\alpha {\bf
s}_\beta -\frac{1}{3}\delta_{\alpha\beta}); \label{ptensor}
\end{eqnarray}
and ${\bf s}$ is the unit vector along the direction of spin
polarization which in our case is along the $z$-axis. It is
evident following the discussions in Ref.(\onlinecite{Zhou04}) that
the vacuum manifold (mean field) of field-induced spin nematic
states is $S^1/Z_2$, identical to that of the standard planar
nematic order parameter.

In the pseudo spin subspace, the above operator can be expressed
in terms of the Pauli matrices $\sigma^{\pm}$,

\begin{eqnarray}
&& \hat{Q}^P_{\alpha\beta}=\sigma^+ \Gamma_{\alpha\beta} +\sigma^- \Gamma^\dagger_{\alpha\beta},
\alpha,\beta=x,y,z;
\end{eqnarray}
and $\Gamma_{\alpha\beta}$ is a constant {\em symmetric traceless} $3\times 3$ matrix.
The detailed form of the $\Gamma$-matrix depends on the number of particles per site  as
well as
the level crossings involved.
One can find more discussions on the calculation of the $\Gamma$-matrix in
Appendix A.

For instance for $2$-particles per lattice site, one finds

\begin{equation}
\Gamma_{\alpha\beta}
=\frac{1}{\sqrt{3}}\left( \begin{array}{ccc}
1 & -i &0\\
-i & -1&0\\
0&0&0
\end{array}\right);
\end{equation}
and for four-particles per site, one obtains

\begin{equation}
\Gamma_{\alpha\beta}
=\sqrt{\frac{14}{15}}\left( \begin{array}{ccc}
1 & -i &0\\
-i & -1&0\\
0&0&0
\end{array}\right).
\end{equation}

It is more convenient to further truncate the projected nematic tensor operator
$\hat{Q}^P_{\alpha\beta}$
in the $xy$-plane which interests us by only keeping the elements with
$\alpha, \beta=x,y$. The truncated operator
then has  the following structure

\begin{eqnarray}
&& \hat{Q}^{Pxy}_{\alpha\beta}=\sigma^+ T_{\alpha\beta} +\sigma^-
T^\dagger_{\alpha\beta},
\alpha,\beta=x,y;
\label{NOPO}
\end{eqnarray}
and $T_{\alpha\beta}$ is a constant {\em symmetric traceless}
$2\times 2$ matrix obtained by truncating the
$\Gamma$-matrix. For $2$-particles per lattice site,

\begin{equation}
T_{\alpha\beta}
=\frac{1}{\sqrt{3}}
\left( \begin{array}{cc}
1 & -i\\
-i & -1
\end{array}\right).
\end{equation}

Finally, the magnetization can also be expressed in terms of the
pseudo spin operators. For instance, the $z$-component is simply
proportional to $\sigma_z$. Following the definition of the pseudo
spin operators, one finds that
\begin{equation}
M_{kz}=\sigma_{kz}+1.
\label{Magnetization}
\end{equation}
(Here we have set the Bohr magneton as one).
The relation between the nematic tensor operator and Pauli
matrices, together with the effective Hamiltonian allow us to
investigate various nematic spin correlations in terms of the
$XXZ$ model.

We also would like to
point out that the subspace spanned by pseudo spin
states is actually a {\em polarization (in the $xy$ plane) free}
subspace. That is

\begin{eqnarray}
&& <{\bf S}_\alpha>=0, \alpha=x,y.
%&& <{\bf S}_\alpha{\bf S}_\beta>-\frac{1}{2} \delta_{\alpha\beta}<{\bf
%S}_\gamma{\bf S}_\gamma>=0
\label{tensor}
\end{eqnarray}
The absence
of polarization (in the $xy$ plane) is a usual property of a
planar spin nematic state.

\section{Critical states as 1D hardcore-boson gases}

As demonstrated in a previous work, the field-driven quantum
phase transitions between spin singlet and spin nematic Mott states
can be studied
in terms of the condensation of spin-two magnons in
spin singlet Mott states\cite{Zhou04}.
To carry out discussions along this line, it is rather convenient to introduce
the Holstein-Primakov representation for the XXZ ferromagnetic spin
model.
In this representation, all spin operators are expressed
in terms of Holstein-Primakov bosons,

\begin{eqnarray}
\sigma^- &=& \left(\sqrt{ 1 - c^\dagger c}\right) c \nonumber \\
\sigma^+ &=& c^\dagger \sqrt{ 1 - c^\dagger c} \nonumber \\
\sigma_z &=& 2 c^\dagger c -1
\label{sigmaz}
\end{eqnarray}
$c^{\dagger}(c)$ is the creation (annihilation) operator of bosons
satisfying the usual bosonic commutation relations $\lbrack c,
c^\dagger \rbrack = 1$; and the raising and lowering operators are
defined as $\sigma^+= ({\sigma_x + i\sigma_y})/{2}$,
$\sigma^-=({\sigma_x - i\sigma_y})/{2}$. One can furthermore
verify that $\lbrack \sigma_\alpha , \sigma_\beta \rbrack = i 2
\epsilon_{\alpha \beta \gamma} \sigma^\gamma$ and ${\bf \sigma}
\cdot {\bf \sigma} =3$. The Hamiltonian of the XXZ model then
transforms into \cite{Zhou04}

\begin{eqnarray}
\frac{{H}_{XXZ}}{\epsilon_0 J_{ex}} &=&
-{2} \sum_{\langle  kl \rangle} \sqrt{\lbrack 1 - c_k^\dagger c_k \rbrack}
c_k
c_l^\dagger \sqrt{1 - c_l^\dagger c_l} \nonumber \\
&&
-{2} \sum_{\langle  kl \rangle} c_k^\dagger \sqrt{(1 - c_k^\dagger c_k)(1
- c_l^\dagger c_l)}
c_l
\nonumber\\
&& - 4 (1 + \beta) \sum_{\langle  kl \rangle } c_k^\dagger c_k c_l^\dagger c_l \nonumber
\\
&& + (4(1+\beta)-2h_z ) \sum_{k} c_k^\dagger c_k;
\label{HPh}
\end{eqnarray}
Previous calculations in Ref.(\onlinecite{Zhou04}) show that
$\beta$ is always negative for interacting spin-one bosons under
consideration.

In the dilute gas limit which interests us,
the number of Holstein-Primakov bosons per lattice site is much less than one i.e.
$n_c=\langle c^{\dagger}_kc_k\rangle  \ll 1$.
For this reason, one can expand the nonlinear operators of $\sigma^{\pm}$
in terms of $n_c$.
Indeed,
the resultant many-body Hamiltonian up to the second order of $n_c$ is

\begin{eqnarray}
&& \frac{{H}_{XXZ}}{\epsilon_0 J_{ex}} = \sum_{q} [4-4 \cos q
+4\beta-2h_z] c_{q}^\dagger c_{q}
\nonumber \\
&& +\frac{1}{V_T}\sum_{ q_1,
q_2,
q} V_B(q) c_{{q}_1+{
q}}^\dagger
c_{{ q}_2-{ q}}^\dagger
c_{{q}_1} c_{{q}_2},\nonumber \\
&& V_B(q)=-{4}\beta \cos q  +O(\rho^2).
\label{int}
\end{eqnarray}
The lattice constant $a$ has been set as one.
And at small $q$ the dimensionless mass of
bosons $m$ is $1/4$.

The Hamiltonian in Eq.(\ref{int}), or Eq.(\ref{HPh}) in fact
describes bosons interacting with short range interactions.
The bare two-body {\em repulsive} interaction potential $V_B(kl)$
when two particles are in two neighboring sites $k$ and $l=k\pm 1$
is $-4 \beta$(in units of $\epsilon_0 J_{ex}$, and $-1< \beta <0$)
i.e.,

\begin{equation}
V_B(kl)=-4\beta \delta_{k\pm 1, l}. \label{vb}
\end{equation}
Therefore, the $q$-dependent two-body interaction constant
becomes $-4 \beta$ as $q$ approaches zero.

Following Eq.(\ref{int}), when $h_z < h_c=2\beta$, all magnon
excitations are fully gapped with the energy gap being $2(
h_c-h_z)$. The ground state therefore is a vacuum of the bosonic
field or a spin singlet Mott state; at $h_z=h_c=2\beta$,
excitations with $q=0$ become gapless signifying a
quantum phase transition between spin singlet states and nematic
states. When $h_z > h_c$, magnons appear in the ground state and
the density is determined by the following standard equilibrium
condition

\begin{equation}
2h_z-2h_c=\mu(\beta, \rho).
\label{equilibrium}
\end{equation}
Here $\mu$ is the chemical potential depending on the density
$\rho$ and the parameter $\beta$. Therefore nematic Mott states
close to critical points are effectively described by dilute
interacting bosons (spinless). Eq.(\ref{equilibrium}) indicates
that close to critical points $h_z-h_c$ approaches zero and bosons
are in the dilute limit.

A similar set of equations have been derived and studied in
Ref.(\onlinecite{Zhou04}). In high-dimensional lattices, one can
easily apply the results on weakly interacting bosons to analyze
field-induced nematic Mott states at and close to critical points.
This has been achieved in a previous work and the relation between
the development of nematic order and {\em condensation} of magnons
has been fully investigated\cite{Zhou04}. The chemical potential
of a weakly interacting magnon condensate with interaction given
in Eq.(\ref{vb}) is simply $-8\beta \rho$. Following
Eq.(\ref{equilibrium}), this leads to the magnon density as well
as the magnetization close to the critical field.

However, in one-dimensional lattices with $V_T$ lattice sites and
a given number of bosons, the single-particle zero point motion
energy scales as $V_T^{-2}$ while the interaction energy per
particle scales as $V_T^{-1}$.
As a result, the dimensionless coupling constant $\gamma$
which is defined as the ratio between the interaction energy per
particle ($-4\rho  \beta$) and the kinetic energy of a particle
confined within a length scale $\rho^{-1}$ ($\rho^2$) is inversely
proportional to $\rho$, the number of particles per
lattice site,

\begin{equation}
\gamma=-\frac{4 \beta }{\rho};
\label{dic}
\end{equation}
and $\gamma$ diverges as $\rho$ approaches zero in the dilute limit close
to critical points.
One-dimensional dilute
bosons with interactions are equivalent to hardcore bosons as the
density $\rho$ is taken to be zero and $\gamma$ becomes infinite.
The critical states in one dimension are distinctly different from free
bosons.

As pointed out by Girardeau\cite{Girardeau60},
the ground state wave functions of hardcore bosons
($\gamma=\infty$) in the continuum limit can be expressed as the
absolute value of the Slater determinant of free fermions.
Liniger and Lieb further investigated cases when $\gamma$ varies
from zero to infinity using the Bethe Ansatz method\cite{Lieb65}.
In addition to confirming the solution in the impenetrable hardcore boson
limit obtained by Girardeau, they also pointed out the important
functional relation between the two-body scattering phase shift
$\theta(q_1,q_2)$ (see more discussions in Appendix B) and the
chemical potential of interacting bosons for an arbitrary
$\gamma$. Especially, at large $\gamma$ limit the chemical
potential of dilute bosons approaches the value of the chemical
potential of free fermions with an identical density.

To understand the development of magnetization
close to critical points, we outline the calculation of the
chemical potential of the Holsetin-Primakov bosons in the dilute
limit. We first notice that in the continuum limit the two-body
interaction in Eq.(\ref{vb}) becomes {\em delta}-like with
strength being equal to $-4\beta$. Following
Ref.\onlinecite{Lieb65} (see more discussions in Appendix B), we
find that the chemical potential for bosons with the two-body
interaction $-4\beta \delta(x-x')$ is determined by the following
set of equations

\begin{eqnarray}
&& \int dq n(q)=\rho, \nonumber\\
&& \frac{N}{\rho}\int dq n(q) \frac{q^2}{2m}=E_0,\nonumber \\
&& -8\beta \int dp \frac{n(p)}{16\beta^2 +(p-q)^2}=2\pi n(q)-1.
\label{Bethe}
\end{eqnarray}
Here $n({q})$ is the distribution function of qausi-momentum; the
dimensionless mass $m$ is equal to $1/4$. $N$ is the total number
of particles. The first two equations above are the definitions of
the number density of particles $\rho$ and the total energy $E_0$
of the system. The last one in Eq.(\ref{Bethe}) follows the Bethe
Ansatz solution. This set of equations have been studied in
Ref.\onlinecite{Lieb65} and for the convenience of readers we
present the derivation in Appendix B. The Bethe Ansatz solution
for the $XXZ$ spin model was also obtained by Yang and
Yang\cite{Yang66}.

The discussion in Appendix B indicates that the solution of
Eq.(\ref{Bethe}) should yield the following chemical potential in
the dilute limit

\begin{equation}
\mu=\frac{\partial E_{0}}{\partial N} =\frac{2
\pi^2}{3}\frac{3\gamma^3+2\gamma^2}{(\gamma+2)^3}\rho^2.
\label{chemical}
\end{equation}
And $\gamma$ is defined in Eq.(\ref{dic}).
In the presence of external fields, in equilibrium
the chemical potential has to be equal to $2h_z-2h_c (>0)$ as suggested
in Eq.(\ref{equilibrium}).
This leads to the
equilibrium density of strongly interacting magnons
close to critical points where the dimensionless interaction constant
$\gamma$ diverges as indicated in Eq.(\ref{dic}),

\begin{equation}
\rho(h_z)=\frac{1}{\pi } \sqrt{h_z-h_c}+...
\label{Bdensity}
\end{equation}
Taking into account Eq.(\ref{Magnetization}),
one obtains

\begin{equation}
M_z=2c^\dagger_k c_k,
\end{equation}
and derives the magnetization close
to the critical point as

\begin{equation}
M_z =\frac{2}{\pi}\sqrt{h_z-h_c}.
\label{Magnetization2}
\end{equation}

We should emphasize here that because of the large dimensionless
coupling constant of dilute bosons in one-dimensional lattices,
the chemical potential is a quadratic function of the density of
bosons instead of a linear function as in the weakly interacting
limit. As a result, in one-dimensional lattices the critical
exponent in the quantum phase transitions between the nematic and
spin singlet Mott states is renormalized to be one-half of the
critical exponent found for high-dimensional lattices (See
Ref.\onlinecite{Zhou04}).

Slightly away from critical points, we also obtain the corrections
to the magnon density as well as the magnetization by expanding
Eq.(\ref{chemical}) in terms of a large but finite $\gamma$. In the
dilute limit, the chemical potential approximately is
\begin{equation}
\mu\simeq\frac{2\pi^2}{3}\left[3\rho^2+\frac{4}{\beta}\rho^3\right].
\end{equation}
The leading correction to the magnetization $M_z$ in
Eq.(\ref{Magnetization2}) is
\begin{equation}
\delta
M_z=\frac{2}{3\pi^2|\beta|}(h_z-h_{c})+\cdot\cdot\cdot
\end{equation}
which becomes significant when $h_z-h_c$ becomes comparable to $\beta^2$.

In Sec. IV, we are going to further examine the critical states using the
Jordan-Wigner Fermionic representation. In Sec. V, we
apply the general bosonization approach for one-dimensional fluid to
the states close to the critical points.
Finally we would like to mention that both bosonic and Majorana fermionic
representation have been employed to study antiferromagnetic spin chains in an
external field\cite{Affleck91,Nomura91,Sorensen93,Tsvelik90}.
Bethe Ansatz solutions of the XXZ model have also been used
to calculate critical exponents in various models\cite{Affleck99}.

\section{Critical states as free-Fermion gases}

Since the ground state wave functions of interacting bosons in the
dilute limit, or equivalently hardcore bosons can be related to
the Slater determinant of free Fermions, not surprisingly many
aspects of the fixed point solutions reminisce free Fermions. It
is certainly appealing to work with the fermionic representation (of
Jordan-Wigner type) and find out to what extend the critical
points can be characterized by the free Fermion model. In this
section we are going to implement this idea.

In the Jordan-Wigner representation, spin-operators
are expressed in terms of string operators defined by Fermions,

\begin{eqnarray}
&& \sigma^+_{k}=
\exp(i\pi \sum_{l<k} f^\dagger_l f_l)
f^\dagger_k \nonumber \\
&& \sigma^-_k=f_k
\exp(-i\pi \sum_{l<k} f^\dagger_l f_l)
\nonumber \\
&& \sigma_{kz}=2f_k^\dagger f_k-1.
\label{JWR}
\end{eqnarray}
The string operator in the Wigner-Jordan form is oriented
along the path starting at $-\infty$ and ending at $k-1$.

Substituting the expressions for spin operators, one rewrites the ferromagnetic XXZ
model in the following form (see more discussions in Appendix C)

\begin{eqnarray}
&& \frac{H_{XXZ}}{\epsilon J_{ex}}=-2\sum_{<kl>}
(f_k^\dagger f_l +f^\dagger_l f_k)+ \nonumber \\
&& + [4 (1+\beta)-2 h_z]\sum_k f_k^\dagger f_k \nonumber \\
&& -{4}(1+\beta)\sum_{<kl>} f_k^\dagger f_k f_l^\dagger f_l.
\label{FermionH}
\end{eqnarray}
Eq.(\ref{FermionH}) shows that the ferromagnetic $XXZ$ model ($0>
\beta >-1$)
is equivalent to {\em attractively} interacting Fermions.
The bare
two-particle attractive interaction $V_F(kl)$
is short ranged; the interaction is nonzero only when two particles occupy
two neighboring sites and has a negative value of $-4(1+\beta)$,i.e.

\begin{equation}
V_F(kl)=-4(1+\beta)\delta_{k\pm 1,l}.
\end{equation}
Obviously the bare potential
vanishes only when $\beta=-1$.

However, because of the Pauli exclusion principle, it is well known
that scattering between spinless Fermions is absent if the
two-particle potential is truly $\delta$- like. For a similar
reason in a dilute Fermi gas the scattering is very weak if the
two-body potential is short ranged. Therefore, in one-dimension
for a given number of Fermions while the one-particle kinetic
energy for Fermions is proportional to $V_T^{-2}$ as in the case
of bosons, the interaction energy per Fermion is proportional to
$V_T^{-3}$ instead of $V_T^{-1}$ as a result of the Pauli
exclusion principle. It indicates that the dilute limit of
Fermions is truly weakly interacting and the fixed point of free
Fermions is stable. Indeed, in one-dimensional lattices, the
dimensionless coupling constant scales as $\rho a$ in the dilute
limit ($a$ is the range of interaction which in our case is the
lattice constant; it is set as one in this article).

Following the discussions in Appendix C, one obtains the following
effective Hamiltonian

\begin{eqnarray}
&& \frac{H_{XXZ}}{\epsilon J_{ex}}=
\sum_q [4 \beta-2 h_z +4-4\cos q ] f_q^\dagger f_q \nonumber \\
&& +\frac{1}{V_T}
\sum_{q_1,q_2,q}
V_F(q)
f^\dagger_{q_1-q}
f_{q_1}
f_{q_2+q}^\dagger f_{q_2};\nonumber \\
&& V_F(q)={4}(1+\beta) (\cos q-1)<0\label{XXZJR-momum}
\end{eqnarray}
%($V_F(q)$ vanishes as $q$ approaches zero. So as a result of the
%Pauli exclusion principle, the forward scattering is suppresed by
%a factor of $q^2$; the backscattering channel is determined by the
%value of $V_F(q)$ at $q=2q_F$ ($q_F$ is the Fermi momentum) and is
%approxmately a constant proportional to $q_F^2$.)
And again at the critical point $h_c=2\beta$, the excitations with
$q=0$ become unstable and the quantum phase transition occurs.
When $h_z < h_c$, Fermionic excitations are gapped with an energy
$2(h_c-h_z)$ and the ground state is a vacuum of Fermions
representing the spin singlet Mott state. Slightly above the
critical field, Fermions with small momenta have negative energies
and we have a dilute interacting Fermi gas in the ground state
representing nematic Mott states.

The second sum in Eq.(\ref{XXZJR-momum}) indicates that the
forward scattering amplitude between right-movers or left-movers
in one-dimension, i.e. $(rq_F, rq_F)$ $\rightarrow$ $(rq_F, rq_F)$
$(r=\pm)$ is proportional to $q^2 \sim 0$ ($q$ denotes the
transferred momentum during the scattering and $q_{F}$ is the
Fermi momentum) and therefore vanishes in the long wave length
limit. However the back scattering between right-movers and
left-movers, $(rq_F, -rq_F)$ $\rightarrow$ $(-rq_F, rq_F)$
involves a momentum transfer $q=2q_F$; the scattering amplitude is
proportional to $q_{F}^2$ and remains finite in the long wave
length limit. As expected before,
the dimensionaless coupling constant which is the ratio
between the interaction energy $\epsilon (1+\beta) \rho $ and the
kinetic energy $\epsilon$ should be proportional to

\begin{equation}
\gamma_F=- (1+\beta)\rho.
\end{equation}

Notice that the bare two-particle attractive interactions
$-4(1+\beta)$ are nonzero
for any system with $\beta\neq -1$ and can even be comparable to the Fermi
energy. In principle, Fermions under consideration can be in the strongly
attracting limit. However, the effective coupling constant
which itself depends on the density of Fermions is always small
in the extremely dilute limit. And it becomes precisely zero when
critical points are approached and one anticipates that the Fermion
density vanishes. Close to the critical points, we do have a collection of weakly
interacting Fermions, a problem which can be easily handled.
Of course when $\beta+1=0$, the bare two-body interactions become
zero identically.

To summarize the result of this mapping, we would like to
emphasize that all states along the triangular boundaries in the phase
diagram in Fig.1 a) correspond to the free Fermions. The $O(3)$
symmetric tricritical point on the other hand is a free boson
critical point. All states within the triangular boundaries are Luttinger
liquids with varying densities and Luttinger liquid parameters.
Finally states in the SSMI phase outside the triangular boundaries in Fig. 1c)
are represented as the vacuums of Jordan-Wigner Fermions and the
SPMI corresponds to a filling factor $N=1$ Fermionic Mott state (see Fig.1 for more discussions).

Therefore, the critical point can also be studied conveniently in the
Fermionic representation. For instance, the chemical potential in this case is
naturally proportional to the square of the Fermion density which is
consistent with the results obtained in the bosonic representation.
Taking into account that $m=1/4$, one obtains

\begin{equation}
\mu_F={2\pi^2\rho^2}
\end{equation}
Along with the equilibrium condition for Fermi gases (which is identical
to the condition for bosons in Eq.(\ref{equilibrium}))

\begin{equation}
\mu_F(\beta,\rho)=2(h_z-h_c),
\end{equation}
one obtains the Fermi density close to the critical point.

Eq.(\ref{Magnetization}) and Eq.(\ref{JWR}) further indicate
that the magnetization is proportional to the Fermion number operator

\begin{equation}
M_{kz}=2 f^\dagger_k f_k.
\end{equation}
This again leads to Eq.(\ref{Magnetization2}).

Moreover, one can express
nematic tensor operator in terms of the Fermion operators

\begin{equation}
\hat{Q}^{Pxy}_{\alpha\beta}=T_{\alpha\beta} \exp(i\pi \sum_{l<k}
f^\dagger_l f_l) f^\dagger_k +h.c.
\end{equation}
In deriving it one has to take into account Eq.(\ref{NOPO})and Eq.(\ref{JWR}), the
mapping between the
pseudo-spin operator and Fermion operators.
Note the nematic operator has nonlocal dependence on the Fermion operators:
it is a string operator of Fermions! This aspect of the tensor operator
plays a very important role in the nematic correlations in 1D.

So we can use the free Fermion properties to calculate the nematic tensor correlators in 1D.
Indeed, we find that the correlator appears
to be clearly related to the free Fermion
correlator,

\begin{eqnarray}
&&
<\hat{Q}^{Pxy}_{\alpha\beta}(k)\hat{Q}^{Pxy}_{\alpha'\beta'}(l)>
=T_{\alpha\beta}T^\dagger_{\alpha'\beta'}
\nonumber \\
&& <f_k \exp(-i\pi \sum_{n<k} f^\dagger_n f_n) \exp(i\pi
\sum_{m<l} f^\dagger_m f_m) f^\dagger_l> +c.c.\nonumber\\
\label{correlator1}
\end{eqnarray}

Evaluating this correlator is tricky because a Fermion operator
itself {\em inexplicitly} contains string operators and is
correlated with the string operator.
The consideration towards the end of Sec. V and
in Appendix C
implies that the contribution from the string operator in the
Jordan-Wigner representation yields precisely a sign oscillation
in the Fermion wave function. So the appearance of the
Jordan-Wigner phase factor in the nematic correlator ensures that
such oscillations in the free Fermion correlator get cancelled
out. Finally, standard calculations yield

\begin{equation}
<\hat{Q}^{Pxy}_{\alpha\beta}(k)\hat{Q}^{Pxy}_{\alpha'\beta'}(l)>
\sim
T_{\alpha\beta}T^\dagger_{\beta^\prime\alpha^\prime}
\frac{1}{|k-l|^{1/\eta_0}},
\label{correlation-fermion}
\end{equation}
where the Friedel oscillations that appear
in the free Fermion correlator are indeed absent in the nematic correlations
(We have also neglected the higher harmonics).
$\eta_0=2$ following the Luttinger liquid theory (see Appendix D).
In the next section, we are going to study the nematic states close to
critical points using the bosonization technique.

\begin{figure}[htbp]
\begin{center}
\includegraphics[width=2.7in]
{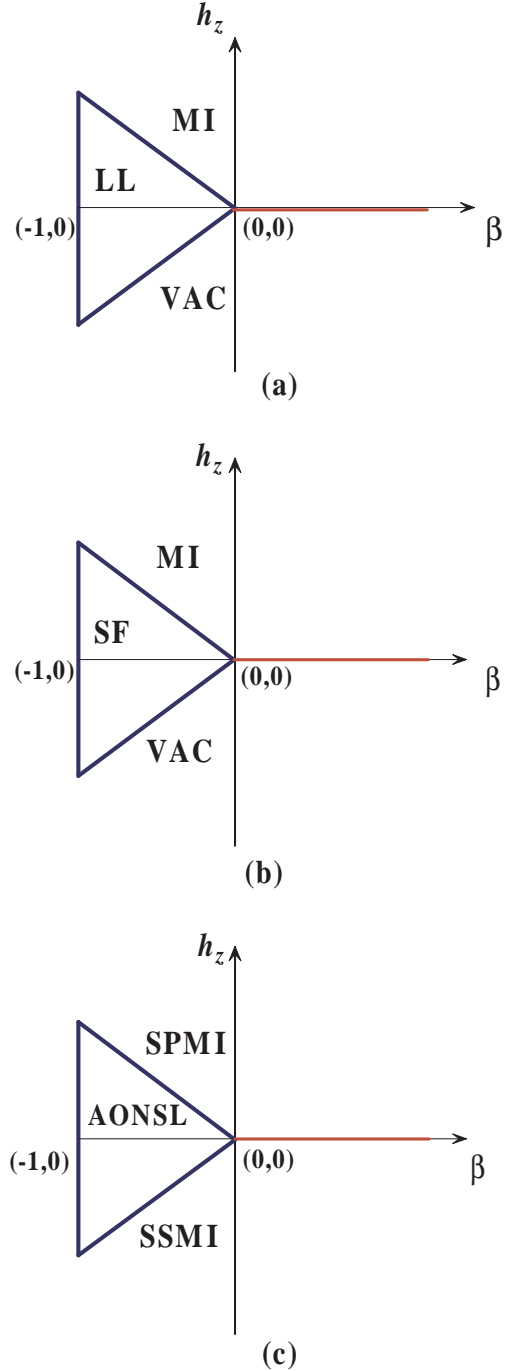}  \caption{\label{phase-diagram} (Color online) a) The
phase diagram for the Jordan-Wigner Fermions in the $h_z$-$\beta$
plane ($\beta \in [-1, \infty]$). Inside the triangle are the
Luttinger liquids (LL). Outside the triangle and $h>0 [<0]$ is the
Mott insulator (MI) [vacuum (VAC)] of spinless Fermions. Along the
boundary of the triangle is the free Fermi gas. b) The phase
diagram for the Holstein-Primakov bosons (1D) in the $h_z$-$\beta$
plane. Inside the triangle is the superfluid (SF). Outside the
triangle and $h>0 [<0]$ is the Mott insulator (MI) [vacuum (VAC)]
of bosons. Along the boundary of the triangle is the hardcore
boson gas. c) The phase diagram for spin-one bosons in 1D
optical lattices indicated by a) and b). Inside the triangle is the
algebraically ordered nematic spin liquid (AONSL). Outside the
triangle and $h>0 [<0]$ is the spin polarized Mott insulator
(SPMI) [spin singlet Mott insulator (SSMI)]. The value of $\beta$
for spin-one bosons was calculated microscopically in
Ref.\onlinecite{Zhou04} and turns out to be always negative. }
\end{center}
\end{figure}

\section{Algebraically ordered spin nematic liquids close to critical points}

Following the discussion at the end of section IV and Appendix C,
as far as the nematic correlations are concerned, we notice that
the Fermion representation of pseudo spin operators is redundant:
the phase string appears twice in the calculation.
Taking this point of view, we find it is more convenient and
natural to employ the bosonic representation (see section III) for
spin operators. And we choose to use the bosonization technique to evaluate
correlations at ($h_z=h_c$) and close to critical points
($h_z>h_c$).

The bosonization scheme is a generic representation to address
interacting bosons as well as Fermions in one-dimension systems.
It is particularly powerful when describing algebraically ordered
quantum spin nematic liquids close to the critical points. And it
also serves very well as a framework to study the
equivalence between the interacting
boson model in section III and the Fermion (free and interacting) models in
section IV; in fact the dynamics of algebraically ordered nematic
spin liquids is identical to Luttinger liquid ones. We are
going to work in the continuum limit.

First we rewrite the Hamiltonian in Eq.(\ref{int}) in terms of
$c^\dagger(x), c(x)$ in the continuum limit ( we have chosen the
lattice constant $a$ to be one and $x$ is dimensionless),

\begin{eqnarray}
&& \frac{H_{XXZ}}{\epsilon J_{ex}}=\int dx c^\dagger (x)
(2h_c-2h_z -\frac{\hbar^2}{2m}{ \partial_x^2})
c(x)\nonumber \\
&&
-4\beta \int dx dx' v_B(x-x') c^\dagger(x) c(x) c^\dagger(x') c(x').
\end{eqnarray}
Here $-4\beta v_B(x-x')$ represents the short ranged two-body potential
in the lattice model and in the continuum limit it is taken to be
a $\delta$-function, i.e., $v_B(x-x')=\delta(x-x')$. The
dimensionless mass of bosons is set to be $m=1/4$.

Following Ref.\onlinecite{Haldane81},
we introduce the following standard conjugating bosonic fields for the
creation operator of bosons $c^\dagger(x)$
(See details in appendix D),

\begin{eqnarray}
&& c^\dagger(x)=\sqrt{\rho(h_z)+\frac{\partial_x
\Theta(x)}{\pi}}
e^{i\Phi(x)}
\sum\limits_{m=-\infty}^{+\infty}e^{2mi\Theta(x)};
\\
&& [\partial_x \Theta(x),\Phi(x')]=i \pi \hbar \delta(x-x').
\label{conjugate}
\end{eqnarray}
And $\rho(h_z)$ has been given in Eq.(\ref{Bdensity}) in section III.

Substituting this Ansatz into Eq.(\ref{HPh}) and taking the long wave
length
limit, we find the following Hamiltonian in terms of $\Theta(x)$ and $\Phi(x)$ ($h_c
>h_z$),

\begin{eqnarray}
&& \frac{H_{XXZ}}{\epsilon J_{ex}}=
(2h_c-2h_z +\mu) \frac{1}{\pi}\int dx \partial_x \Theta(x) +\nonumber \\
&&  \int dx \left[\frac{1}{2\pi^2}\frac{\partial \mu}{\partial
\rho} (\partial_x \Theta(x))^2  + \frac{\hbar^2}{2m}{\rho}(h_z)
(\partial_x \Phi(x))^2 \right]. \label{bosonization}
\end{eqnarray}

For an unrenormalized theory, the {\em bare} value of $\mu(\beta,
h_z)$ is taken to be $-8\beta \rho(h_z)$ and $\partial
\mu/\partial \rho=-8\beta$. Since the theory is renormalizable at
the long wave length limit, we expect that the low energy
effective Hamiltonian should be of the same form as in the bare
case except that quantities such as $\mu$ can be strongly
renormalized by short wave length interactions. In fact this form
of the effective Hamiltonian is suggested by the
hydrodynamics of bosonic gases\cite{Haldane81,Nozieres90}, though the
$\gamma$-dependence of the renormalized $\mu$ and $\partial
\mu/\partial \rho$ can be obtained only when certain microscopics
are taken into account.

Therefore,
the effective Hamiltonian only depends on the density and the
renormalized functional form of the chemical
potential $\mu(\rho)$. And both the chemical potential and the
compressibility
$\partial \mu/\partial \rho$ are functions of the dimensionless
interaction constant $\gamma (\beta, h_z)$ defined in Eq.(\ref{dic}).
Following Eq.(\ref{Bdensity}), one finds that

\begin{equation}
\gamma(\beta, h_z)=\frac{4 \pi \beta}{\sqrt{h_z - h_c}}
\label{dic1}
\end{equation}
as a function of $h_z$.

In equilibrium, the first term in Eq.(\ref{bosonization}) which is
linear in $\Theta(x)$ has to vanish because of the stability of
the system. This sets the known relation between the chemical
potential and external field, $2(h_z-h_c)=\mu(\rho)$ which has
been discussed in the previous sections. For the weakly
interacting systems, in the leading order of $\gamma$, the
chemical potential takes the {\em bare} value of $-8\beta \rho$;
for the strongly interacting case as in the 1D dilute limit, $\mu$
is renormalized and discussions in section III show that $\mu$ is
proportional to $2\pi^2 \rho^2$ in the leading order of $1/\gamma$.

Taking this into account, we substitute into Eq.(\ref{bosonization})
the renormalized chemical potential $\mu$. We recast it in a more
familiar form

\begin{eqnarray}
&& \frac{H_{XXZ}}{\epsilon J_{ex}}= \frac{\hbar}{2 \pi} \int dx
\left[v_N (\partial_x \Theta(x))^2 + v_J (\partial_x \Phi(x))^2
\right]. \nonumber\\\label{bosonization1}
\end{eqnarray}
Here
\begin{equation}
v_N(\beta, h_z)=\frac{1}{\pi \hbar} \frac{\partial \mu(\rho)}{\partial
\rho(h_z)},
v_J= \pi \hbar \frac{\rho(h_z)}{m};
\end{equation}
again $m=1/4$.
An alternative way of calculating $v_N$ can be found in
an early work\cite{Haldane81}. $v_J$ here is taken to be independent of
interactions as a result of Galilean invariance (approximate) in the continuum limit;
in general when periodical or random potentials are present, $v_J$ could be
renormalized away from this value.

To calculate $v_N$ and $v_J$, one has to take into account the
Bethe Ansatz result for the chemical potential in
Eq.(\ref{chemical}). One obtains the following results for $v_N$,
\begin{equation}
v_{N}=4\pi\hbar\rho(h_{z})\frac{(-4\beta)^4}{[-4\beta+2\rho(h_z)]^4}
\end{equation}
Eq.(\ref{Bdensity}) shows that both $v_N$ and $v_J$ vanish as $h_z$
approaches the critical field $h_c$. The discussions in
Appendix D show that the nematic operator is algebraically
correlated, and close to the critical point it takes the form

\begin{equation}
\langle
Q_{\alpha\beta}^{P_{xy}}(k)Q_{\alpha^\prime\beta^\prime}^{P_{xy}}(l)\rangle\sim
\frac{T_{\alpha\beta}T_{\alpha^\prime\beta^\prime}(1-\rho)\rho}
{(\rho|k-l|)^{1/\eta}}.
\end{equation}
in the long-wave length limit, where the correlation exponent
is determined by the value of $\eta$
\begin{equation}
\eta(\beta, h_z)=2\sqrt{\frac{v_J}{v_N}}=
2{\left[1+\frac{\sqrt{h_{z}-h_{c}}}{2\pi\beta}\right]^2}.
\label{vN}
\end{equation}
In Fig.1 one can find more discussions on various phases in the
bosonic representation and implications on spin-one bosons in
one-dimensional lattices.

The sound velocity of the Holstein-Primakov bosons
corresponds to the spin-wave velocity in nematic states.
Close to the critical points when
$\rho(h_z)$ vanishes, $v_N\approx v_J$ and the spin wave velocity
$v_s=\sqrt{v_N v_J}$ is

\begin{equation}
v_s (h_z) \approx 4\pi\hbar\rho(h_z)=4\hbar\sqrt{h_z-h_c}+...
\end{equation}
following Eq.(\ref{Bdensity}).

Eq.(\ref{conjugate}) and Eq.(\ref{bosonization1}) also define the
Luttinger
liquid dynamics of interacting Fermions\cite{Luttinger63,Mattis65,Tomonaga50,Luther74}.
In the standard formulation of the Luttinger liquid theory\cite{Luther74},
the left-moving and
right moving
Fermion
operators are expressed in terms
of bosonic fields of $\Theta(x)$ and $\Phi(x)$

\begin{equation}
f^\dagger_{L,R}=\exp(\pm i q_F x)
\exp(i\Theta(x)\pm i\Phi(x)).
\label{LRMOVER}
\end{equation}
The dynamics of $\Theta(x)$ and $\Phi(x)$ is given by

\begin{eqnarray}
&& H_{L.L.}=\frac{\hbar \sqrt{v_N v_J}}{2}\int dx [\frac{K}{\pi} (\partial_x
\Phi(x))^2
+\frac{\pi}{K} \Pi^2(x)], \nonumber \\
\label{HLL}
\\
&& [\Pi(x),\Phi(x')]=i\hbar \delta(x-x').
\label{conjugateLL}
\end{eqnarray}
Here $\Pi(x)=\frac{1}{\pi}\partial \Theta(x)$.
One can easily verify that Eq.(\ref{bosonization1}) and
Eq.(\ref{conjugate}) are identical to Eq.(\ref{HLL})
and Eq.(\ref{conjugateLL}) if one identifies that

\begin{equation}
\eta=2K=2\sqrt{\frac{v_J}{v_N}}.
\end{equation}
This intimate relation between 1D bosonic superfluid and Luttinger liquids
was appreciated before\cite{Haldane81}.

\section{Conclusions and Discussions}

In this article, we have studied the critical phenomenon in the
field-driven quantum phase transitions between the nematic Mott
state and spin singlet Mott state in one-dimensional lattices. The
critical states in one-dimension are in the universality class of
impenetrable bosons or free Fermions. We have found that because
of hardcore nature of magnons in one-dimension the critical
exponent of magnetization is renormalized to be one-half of the
value in three-dimensional lattices.

We have also calculated the nematic spin correlations using the
Bosonization scheme and obtained Zeeman-field-dependent algebraically
ordered nematic spin liquids.
At critical points, the nematic spin correlation functions are
identical to two-point correlations of one-dimensional free Fermions
with the Friedel oscillations completely suppressed.
Close to critical points, the algebraically nematically ordered
states are equivalent to
Luttinger liquids of weakly attractive spinless Fermions or
fluid of strongly repulsive bosons.

F.Z. thanks CASTU, Tsinghua University, Beijing for its
hospitality during his stay in the summer of year 2004;
he would like to acknowledge useful discussions with I. Affleck. This
project is in part supported by NSF, P. R. China(Grant No. 10247002 and
10404015)(H.Z.) and in part supported by a grant from UBC and NSERC, Canada
(F.Z.).

Note added: The transition described here might be considered as an
analogue of commensurate- incommensurate transitions and
is of quantum Pokrovsky- Talapov type\cite{Pokrovsky79}.

\appendix

\section{Nematic tensor order parameter in the pseudo spin subspace}

Close to critical points, the Hilbert space is truncated into a
two-level subspace at each site. Therefore we can define a
pseudo-spin $1/2$ at each site, where $|\uparrow\rangle$ denotes
the state $|S+2,S_{z}=S+2\rangle$ and $|\downarrow\rangle$ denotes
the state $|S,S_{z}=S\rangle$. In this truncated subspace the
nematic operator
\begin{equation}
\hat{Q}_{\alpha\beta}=\psi^\dag_{\alpha}\psi_{\beta}-\frac{1}{3}\delta_{\alpha\beta}
\psi^\dag_{\gamma}\psi_{\gamma},
\end{equation}
can be expressed in terms of pseudo-spin operators.

Consider the case of $2$-particle per site. After some
straightforward calculations, the matrix elements of the nematic
operator can be obtained as 
\begin{equation}
\langle\uparrow|\hat{Q}_{\alpha\beta}|\uparrow\rangle=\left(\begin{array}{ccc}
\frac{1}{3} & i & 0 \\
-i  &  \frac{1}{3}&
 0 \\
 0 & 0 & -\frac{2}{3}
\end{array}  \right),
\end{equation}
\begin{equation}
\langle\uparrow|\hat{Q}_{\alpha\beta}|\downarrow\rangle=\frac{1}{\sqrt{3}}\left(\begin{array}{ccc}
1 & -i & 0 \\
-i  &  -1 & 0 \\
 0 & 0 & 0
\end{array}
\right)=\langle\downarrow|\hat{Q}_{\alpha\beta}|\uparrow\rangle^*,
\label{Qmatrix}
\end{equation}
and
\begin{equation}
\langle\downarrow|\hat{Q}_{\alpha\beta}|\downarrow\rangle=0.
\end{equation}
Therefore, for an arbitrary state $|\Omega\rangle$ in this
subspace
\begin{equation}
|\Omega\rangle=\cos\frac{\theta}{2}\exp{\left(-i\frac{\phi}{2}\right)}|\uparrow\rangle
+\sin\frac{\theta}{2}\exp{\left(i\frac{\phi}{2}\right)}|\downarrow\rangle,
\end{equation}
the expectation value of the nematic operator is
\begin{eqnarray}
\langle\Omega|\hat{Q}_{\alpha\beta}|\Omega\rangle=&&\cos^2\frac{\theta}{2}\left(\begin{array}{ccc}
\frac{1}{3} & i & 0 \\
-i  &  \frac{1}{3} & 0 \\
 0 & 0 & -\frac{2}{3}%
\end{array}  \right)\nonumber \\+&&\sin\theta\exp{(i\phi)}\frac{1}{2\sqrt{3}}
\left(\begin{array}{ccc}
1 & -i & 0 \\
-i  &  -1 & 0 \\
0 & 0 &  0%
\end{array}  \right)\nonumber \\+&&\sin\theta\exp{(-i\phi)}\frac{1}{2\sqrt{3}}
\left(\begin{array}{ccc}
1 & i & 0 \\
i  &  -1 & 0 \\
0 & 0 & 0%
\end{array}  \right).
\end{eqnarray}
The first term in the right-hand side of above equation is
contributed from the fully polarized state and should be projected
away. The remained two terms give the expectation value of the
projected nematic operator, which is
\begin{eqnarray}
\langle\Omega|\hat{Q}_{\alpha\beta}^{P}|\Omega\rangle=\sin\theta\exp{(i\phi)}\frac{1}{2\sqrt{3}}
\left(\begin{array}{ccc}
1 & -i & 0 \\
-i  &  -1 & 0 \\
0 & 0 &  0%
\end{array}  \right)\nonumber\\+\sin\theta\exp{(-i\phi)}\frac{1}{2\sqrt{3}}
\left(\begin{array}{ccc}
1 & i & 0 \\
i  &  -1 & 0 \\
0 & 0 & 0%
\end{array}  \right).\label{Expectation of Q}
\end{eqnarray}

To express the projected nematic operator in terms of
pseudo-spin operators, one should look for a suitable combination
of the $\sigma$-matrices which yields the same expectation value
as Eq.(\ref{Qmatrix}), or Eq.(\ref{Expectation of Q}).
It is easily to verify that $\hat{Q}^P_{\alpha\beta}$ should be
represented by the
following operator in the pseudo-spin space,
\begin{equation}
\frac{1}{2}(\sigma_{x}+\sigma_{y})\Gamma_{\alpha\beta}+\frac{1}{2}(\sigma_{x}-i\sigma_{y})\Gamma^{*}_{\alpha\beta}
,
\end{equation}
where
\begin{equation}
\Gamma_{\alpha\beta} =\frac{1}{\sqrt{3}}\left( \begin{array}{ccc}
1 & -i &0\\
-i & -1&0\\
0&0&0
\end{array}\right).
\end{equation}

\section{The chemical potential of interacting bosons in one-dimension}

The one-dimensional interacting bosons problem can be exactly
solved by the Bethe Ansatz method. In this appendix we will
briefly review the Betha Ansatz method used in
Ref.\onlinecite{Lieb65}. Let $\{q_{i}\}(i=1,\cdot\cdot\cdot,N)$ be
the quasi-momenta of the $N$ bosons, the Bethe Ansatz solution
indicates that all the allowed values of $q_{i}$'s should satisfy
the following set of equations:
\begin{equation}
(-1)^{N-1}e^{-iq_{j}L}=\exp\left(i\sum\limits_{i=1}^{N}\theta(q_{i}-q_{j})\right),
\end{equation}
where $\theta_{i,j}$ is the phase shift. In our case, one finds that
\begin{equation}
\theta(q_{i}-q_{j})=-2\arctan\frac{q_{i}-q_{j}}{-4\beta}.
\end{equation}
Dividing two successive equations and equating exponents,
one obtains
\begin{equation}
(q_{i+1}-q_{i})L=\sum\limits_{j=1}^{N}(\theta_{j,i}-\theta_{j,i+1})+2\pi
I_{i}\label{Bethe-equation},
\end{equation}
where a set of integers $I_{i}(i=1,\cdot\cdot\cdot, N)$ are the
quantum numbers for this system, and they are all equal to one for
the ground state. Taking the continuum limit of
Eq.(\ref{Bethe-equation}), one has
\begin{equation}
-8\beta\int\limits_{-\Lambda}^{\Lambda}dp\frac{n(p)}{16\beta^2+(p-q)^2}=2\pi
n(q)-1,
\end{equation}
where $\Lambda$ is the cut-off momentum which should be determined
self-consistently. Here $n(q)$ is the distribution function of
quasi-momentum, which satisfies the constraint
\begin{equation}
\int\limits_{-\Lambda}^{\Lambda}dqn(q)=\rho.
\end{equation}
And the ground state energy is given by
\begin{equation}
E_{0}=\frac{N}{\rho}\int\limits_{-\Lambda}^{\Lambda}n(q)\frac{q^2}{2m}dq.
\end{equation}

Now rescale all qualities in terms of the cut-off momentum $\Lambda$,
i.e. $-4\beta=\Lambda\lambda$, $p=\Lambda y$ and $q=\Lambda x$,
and denote $n(q)$ as $g(x)$. The above three
equations take the following form
\begin{equation}
1+2\lambda\int\limits_{-1}^{1}\frac{g(y)}{\lambda^2+(y-x)^2}dy=2\pi
g(y)\label{integral-equation},
\end{equation}
\begin{equation}
\gamma\int\limits_{-1}^{1}g(x)dx=\lambda \label{constraint},
\end{equation}
and
\begin{equation}
E_{0}=\frac{1}{2m}N\rho^2\frac{\gamma^3}{\lambda^3}\int\limits_{-1}^{1}g(x)x^2dx\label{Genergy}.
\end{equation}

In the limit of large $\gamma$, $\lambda$ is large compared to
$|x-y|$, one can then neglect the $x$ dependence of the integrand
in Eq.(\ref{integral-equation}); and $g(x)$ turns out to be a
constant. The integral equation can be solved approximately and
the result is
\begin{equation}
g=\frac{\lambda}{2\pi\lambda-4}.
\end{equation}
Substituting it back to Eq.(\ref{constraint}) one obtains the
relation between $\lambda$ and $\gamma$ as
\begin{equation}
\lambda=\frac{\gamma+2}{\pi}.\label{lambda-gamma2}
\end{equation}
And finally, using Eq.(\ref{Genergy}) one can derive the ground
state energy
\begin{equation}
E_{0}=N\rho^2\frac{\pi^2}{6m}\left(\frac{\gamma}{\gamma+2}\right)^2.\label{e-gamma2}
\end{equation}
Consequently, a relation between chemical potentials and particle
densities is established in this limit,
\begin{equation}
\mu=\frac{\partial E_{0}}{\partial
N}=\frac{\pi^2}{6m}\frac{3\gamma^3+2\gamma^2}{(\gamma+2)^3}\rho^2.
\end{equation}

\section{The Jordan-Wigner Fermion representation}

\subsection{The effective Hamiltonian for Jordan-Wigner Fermions in the
XXZ model}

In this appendix we will briefly review the Jordan-Wigner
Fermionic representation of the XXZ model, the Hamiltonian of
which is
\begin{eqnarray}
\frac{H_{XXZ}}{\epsilon
J_{\text{ex}}}=-2\sum\limits_{<kl>}\left(\sigma^{+}_{k}\sigma^{-}_{l}+\sigma^{-}_{k}\sigma^{+}_{l}
\right)\nonumber\\-(\beta+1)\sum\limits_{<kl>}
\sigma_{kz}\sigma_{lz}-h_{z}\sum\limits_{k}\sigma_{z}\label{XXZJR}.
\end{eqnarray}
Following the Jordan-Wigner transformation Eq.(\ref{JWR}), the spin
flipping term becomes
\begin{eqnarray}
&&\sigma^{+}_{k}\sigma^{-}_{k+1}=e^{i\pi\sum\limits_{l=1}^{k-1}f^\dag_{l}f_{l}}f^\dag_{k}f_{k+1}
e^{-i\pi\sum\limits_{l=1}^{k}f^\dag_{l}f_{l}}=f^\dag_{k}e^{-i\pi
f^\dag_{k}f_{k}}f_{k+1}\nonumber\\
&&=f^\dag_{k}(1-2f^\dag_{k}f_{k})f_{k+1}=f^\dag_{k}f_{k+1},
\end{eqnarray}
and similarly,
\begin{eqnarray}
\sigma^{+}_{k+1}\sigma^{-}_{k}=f^\dag_{k+1}f_{k}.
\end{eqnarray}
These yield a kinetic term of these Fermions. Neglecting a
constant term, the second term in Eq.(\ref{XXZJR}) turns out to be
\begin{equation}
\sum\limits_{kl}\sigma_{kz}\sigma_{lz}
=4\sum\limits_{kl}f^\dag_{k}f_{k}f^\dag_{l}f_{l}-4\sum\limits_{k}f^\dag_{k}f_{k}.
\end{equation}
Thus one arrives at the Jordan-Wigner Fermion representation of
the XXZ model as shown in Eq.(\ref{FermionH}).

Finally,
due to the Pauli exclusive principle one can treat the interacting terms
as follows
\begin{eqnarray}
&&\sum\limits_{k}f^\dag_{k}f^\dag_{k+1}f_{k}f_{k+1}=\sum\limits_{k}\left(f^\dag_{k}f^\dag_{k+1}
f_{k}f_{k+1}-f^\dag_{k}f^\dag_{k}f_{k}f_{k}\right)\nonumber\\
&&=\sum\limits_{q,q_{1},q_{2}}f^\dag_{q_{1}-q}
f^\dag_{q_{2}+q}f_{q_{1}}f_{q_{2}}(\cos q-1),
\end{eqnarray}
and this results in Eq.(\ref{XXZJR-momum}).

\subsection{Calculation of Eq.(\ref{correlation-fermion})}

In the continuum limit, $\sigma^{\pm}$ can be written in the
following form

\begin{eqnarray}
&& \sigma^+=\exp(-i q_F x -i\Theta(x)) f^\dagger(x);\nonumber \\
&& \sigma^-=f(x) \exp(iq_F x +i\Theta(x)).
\end{eqnarray}
Taking into account the expression for the left- and right-movers
in Eq.(\ref{LRMOVER}),
one obtains

\begin{eqnarray}
&& \sigma^+\sim \exp(i\Phi(x))+\exp(-i2 q_F x-i\Phi(x));\nonumber \\
&& \sigma^-\sim \exp(-i\Phi(x))+\exp(i 2q_F x+i\Phi(x)).
\end{eqnarray}

Therefore the leading contribution to the lowest harmonics
in the correlation function is

\begin{equation}
<\sigma^+\sigma^-> \sim  <\exp(i\Phi(x)-i\Phi(0))>.
\end{equation}
Calculation of this correlator is the same as those in Appendix D.
For free Fermions, one obtains the result in
Eq.(\ref{correlation-fermion}).

\section{Bosonization and nematic spin correlations}

In this appendix we briefly review the bosonization procedure
for one-dimensional interacting bosonic
system\cite{Haldane81}. First we express the bosonic
operator as
\begin{equation}
c^\dag(x)=\sqrt{\rho_T(x)}e^{i\Phi(x)}.
\end{equation}
The mean ground state
density is denoted by $\rho_{0}$, we introduce a field $\Pi(x)$ to
account for the long-wave length ($>\rho_{0}^{-1}$) density
fluctuations. Furthermore in order to take the discreteness of
particles into account, we introduce an auxiliary field
$\Theta(x)$ which is related to $\Pi(x)$ via
\begin{equation}
\frac{1}{\pi}\partial_{x}\Theta(x)=\rho_{0}+\Pi(x),
\end{equation}
where $\Theta$ increases monotonically by $\pi$ each time when $x$
passes the location of a particle. Therefore the total density can
be expressed as
\begin{equation}
\rho_T(x)=\frac{1}{\pi}\partial_{x}\Theta(x)\sum\limits_{n=-\infty}^{+\infty}\delta(\Theta(x)-n\pi),
\end{equation}
where the discrete character is
incorporated. Since the square root of a $\delta$ function is also
proportional to a $\delta$ function, it yields that
\begin{eqnarray}
c^\dag(x)&\sim&
\sqrt{\rho_{0}+\Pi(x)}\sum\limits_{n=-\infty}^{+\infty}\delta(\Theta(x)-n\pi)e^{i\Phi(x)}\nonumber\\
&=&\sqrt{\rho_{0}+\Pi(x)}
\sum\limits_{m=-\infty}^{+\infty}e^{2mi\Theta(x)}e^{i\Phi(x)},
\end{eqnarray}
where the second equality follows from the Poison's summation
identity
\begin{equation}
\sum\limits_{n=-\infty}^{+\infty}f(n)=\sum\limits_{m=-\infty}^{+\infty}\int_{-\infty}
^{\infty}dz f(z)e^{2m\pi iz}.
\end{equation}
It can be verified that the field $\Theta$ and $\Phi$ satisfy the
commutative relation
\begin{equation}
[\partial_{x}\Theta(x),\Phi(x^\prime)]=i\pi\hbar\delta(x-x^\prime).
\end{equation}

With these expressions, the effective Luttinger Hamiltonian can be
read as
\begin{eqnarray}
H_{\text{eff}}=\frac{\hbar}{2\pi}\int_{0}^{L}dx[v_{J}(\partial_{x}\Phi(x))^2+v_{N}
(\partial_{x}\Theta(x))^2].
\end{eqnarray}
Here we have introduced two phenomenological parameters $v_{J}$
and $v_{N}$. $v_{J}$ is the phase stiffness which equals to
$\hbar\pi\rho/m$ in a uniform system, and $v_{N}$ is the
density stiffness which is defined as
\begin{equation}
v_{N}=\frac{1}{\pi \hbar}\left(\frac{\partial \mu}{\partial
\rho}\right).\label{vN1}
\end{equation}
These parameters should be determined by microscopic calculations.
For boson gases with delta-like interactions, these parameters can
be obtained from the Bethe Ansatz solutions.

Furthermore we can express the correlation of nematic operator
in terms of correlators of Holstein-Primakov bosons,

\begin{eqnarray}
&&\langle
Q_{\alpha\beta}^{P_{xy}}(k)Q_{\alpha^\prime\beta^\prime}^{P_{xy}}(l)\rangle
=T_{\alpha\beta}T^\dag_{\beta^\prime
\alpha^\prime}\nonumber\\&&\langle
c^\dag(k)\sqrt{1-c^\dag(k)c(k)}\sqrt{1-c^\dag(l)c(l)}c^\dag(l)\rangle+h.c.\nonumber\\
&&\simeq T_{\alpha\beta}T^\dag_{\beta^\prime
\alpha^\prime}(1-\rho)\langle c^\dag(k)c(l)\rangle+h.c.
\end{eqnarray}
The correlation function for the Holstein-Primakov bosons can be
computed using the effective Luttinger Hamiltonian,

\begin{eqnarray}
&&\langle c^\dag(x)c(0)\rangle=\rho\left[\frac{1}{\rho|x|}
\right]^{1/(2K)}\left\{ b_{0}\right. \nonumber\\
&&\left.+\sum\limits_{m=1}^{+\infty}b_{m}\left(\frac{1}{\rho|x|}
\right)^{2m^2K}\cos(2\pi m\rho x)\right\}.
\end{eqnarray}
Here
\begin{equation}
\eta=2K=2\sqrt{\frac{v_{J}}{v_{N}}}
\end{equation}
is a universal parameter only depending on long-wave length
physics, while the $b_{m}$'s depend on the microscopic details.

%\end{multicols}
\end{document}